\documentclass[preprint]{aastex}

\newcommand\arcpt{${{\lower3pt\hbox{$^{\prime\prime}$}}\atop{\raise4pt\hbox{.}}}$}

\slugcomment{To appear in the April 2002 issue of the {\it Astronomical Journal}}

\shorttitle{Henry et al.}
\shortauthors{New Nearby Southern Stars}

\begin{document}

\title{The Solar Neighborhood VI: \\
       New Southern Nearby Stars Identified by Optical Spectroscopy}

\author{Todd J. Henry\altaffilmark{1}}
\affil {Department of Physics and Astronomy, Georgia State University,
        Atlanta, GA 30303}

\author{Lucianne M. Walkowicz}
\affil {Department of Physics and Astronomy, Johns Hopkins University,
        Baltimore, MD 21218}

\author{Todd C. Barto\altaffilmark{1}}
\affil {Lockheed Martin Aeronautics Company, Boulder, CO 80306}

\and

\author{David A. Golimowski}
\affil {Department of Physics and Astronomy, Johns Hopkins University, 
        Baltimore, MD 21218}

\altaffiltext{1}{Visiting Astronomer, Cerro Tololo Inter-American
Observatory.  CTIO is operated by AURA, Inc.\ under contract to the
National Science Foundation.}


\begin{abstract}

Broadband optical spectra are presented for 34 known and candidate
nearby stars in the southern sky.  Spectral types are determined using
a new method that compares the entire spectrum with spectra of more
than 100 standard stars.  We estimate distances to 13 candidate nearby
stars using our spectra and new or published photometry.  Six of these
stars are probably within 25~pc, and two are likely to be within the
RECONS horizon of 10 pc.

\end{abstract}


\clearpage


\section{Introduction}

The nearest stars have received renewed scrutiny because of their
importance to fundamental astrophysics ({\it e.g.}, stellar
atmospheres, the mass content of the Galaxy) and because of their
potential for harboring planetary systems and life ({\it e.g.}, the
NASA Origins and Astrobiology initiatives).  The smallest stars, the M
dwarfs, account for at least 70\% of all stars in the solar
neighborhood and make up nearly half of the Galaxy's total stellar
mass (Henry {\it et al.} 1997; hereafter H97).  Their slightly lesser
cousins, the brown dwarfs, may lurk in comparable numbers.  Yet, many
of the nearest red, brown, and white dwarfs remain unrecognized
because of their low luminosities.  H97 estimate that more than 30\%
of stellar systems within 10~pc of the Sun are currently missing from
compendia of nearby stars.

The number of ``missing'' stars within 25~pc of the Sun is estimated
to be twice the fraction missing within 10~pc.  The NASA/NSF NStars
Project is a new effort to foster research on all stars within 25~pc,
with special emphasis on the development of a comprehensive NStars
Database.  All systems with trigonometric parallaxes greater than or
equal to 0\arcpt04000 from {\it The Yale Catalog of Stellar
Parallaxes} (YPC, van Altena {\it et al.} 1995) and {\it The Hipparcos
Catalogue} (HIP, ESA 1997) have been included in the Database.  The
weighted means of the YPC and HIP parallaxes have been determined,
including the combination of all trigonometric parallax values for
stellar systems in which widely separated components have had separate
parallax measurements.  Table~1 lists the numbers of known and
predicted stellar systems within 25~pc, and their distributions within
equal regions of the sky, obtained from the NStars Database as of 01
July 2001.  The predicted number of 1375 systems in each region is
based on the assumptions that (1) the density of stellar systems
within 5~pc (0.084 systems~pc$^{-3}$) extends to 25~pc, and (2) the
distribution of the systems is isotropic.  Table~1 clearly shows that
more stars are missing in the southern sky than in the northern sky
--- we predict that more than two-thirds of the systems are
undiscovered in the south.  Furthermore, new systems within 5~pc are
still being found (H97), so the total number predicted within 25~pc is
a lower limit.

In a concerted effort to discover and characterize the nearest stars,
the RECONS (Research Consortium on Nearby Stars) team has been
conducting astrometric, photometric, spectroscopic, and multiplicity
surveys of known and candidate stars within 10~pc (for more
information about RECONS, see H97).  In this paper, the sixth in {\it
The Solar Neighborhood} series, we present optical spectra of 34 known
or suspected nearby southern red and white dwarfs, including 10 known
members of the RECONS sample and 16 stars for which no spectral types
have been previously published.  We report spectral types for all the
stars in our sample using a method that will define the spectral types
used in the NStars Database.  We supplement the spectral data with VRI
photometry for five stars.  Our analysis has revealed two new stars
that are probably closer than the 10~pc RECONS horizon, and four
others that are probably closer than the 25~pc NStars horizon.


\section{Sample}

The 34 stars for which we obtained optical spectra are grouped into four 
categories:

\begin{enumerate}

\item {\it 12 stars that lie within, or close to, the 10~pc RECONS
horizon for which no broadband spectra are published.}  These stars
have well known distances, so they are good standards for calibrating
spectroscopic parallaxes.

\item {\it 14 recently discovered stars having high proper motions.}
Because nearer stars appear to move faster than more distant ones,
high proper motion is a good indicator of stars in the solar
neighborhood.  Between 1989 and 1997, Wroblewski and collaborators
identified 2055 new stars lying south of $-5^{\circ}$ declination with
proper motions, $\mu$ $\ge$ 0\arcpt15~yr$^{-1}$ (Wroblewski \& Torres
1997, and references therein).  In February 1998, we observed 12 of
the 52 stars from this collection that have
$\mu~\ge~0$\arcpt50~yr$^{-1}$.  We also observed two
high-proper-motion stars selected from the Calan-ESO survey of Ruiz
and collaborators (Ruiz {\it et al.} 1993).  Between our observing run
and the end of 2000, Wroblewski and collaborators identified an
additional 293 new stars with $\mu~\ge~0$\arcpt15~yr$^{-1}$, only one
of which has $\mu~\ge~0$\arcpt50~yr$^{-1}$ (Wroblewski \& Costa 1999).

\item {\it 4 stars whose Hipparcos parallaxes have suspiciously high
errors.}  Nine targets in eight systems were reported by the Hipparcos
mission to have parallaxes larger than 0\arcpt100 with errors larger
than 0\arcpt020 ({\it i.e.}, 14--56\% errors -- enormous for
Hipparcos).  In every case, the targets are near bright stars that
corrupted the parallax measurements.  In two cases, HIP~114110 and
HIP~114176, there is no star at all.  In February 1998, we observed
two of the remaining seven targets, HIP~15689 and HIP~20698, as well
as two of the neighboring bright stars.

\item {\it 4 stars for which available photometry implies a distance
less than 25~pc.}  These stars come from an extensive list of possible
nearby stars maintained by the first author.

\end{enumerate}


\section{Observations}

A total of 34 targets were observed during nights of UT 08 and 09
February 1998 using the Blanco 4m telescope at Cerro Tololo
Interamerican Observatory (CTIO).  The R-C Spectrograph with a Loral
3K~$\times$~1K CCD was used with grating \#181 at tilt
58.77$^{\circ}$, order blocking filter OG-515, and a gain setting of 4
($2.07~e^-$~ADU$^{-1}$).  The wavelength coverage was 5500--10000~\AA,
with a resolution of 6~\AA.  This broad spectral range includes the
TiO absorption bands characteristic of red dwarfs, the H$\alpha$
emission line used to measure activity, and the K~I, Na~I, and Ca~II
absorption features used to discriminate dwarfs from giants.

Bias frames and dome flats were taken at the beginning of each night.
A HeAr arc lamp spectrum was recorded after each target spectrum to
permit accurate dispersion correction throughout the night.
Observations were conducted through clouds for four hours on 08
February 1998, and for eight hours through increasing clouds on 09
February 1998.  Several of the program stars were observed on both
nights, thereby allowing confirmation of the spectral types.

Photometric observations were carried out in the $V_J$, $R_C$, $I_C$
bands for five stars at the CTIO 0.9m telescope during our NOAO
Surveys Program, CTIOPI (CTIO Parallax Investigation), on the nights
of UT 27 November to 01 December 1999.  Standards from Landolt (1992)
and Bessell (1990) were observed for the purpose of deriving
extinction coefficients and transformation equations for each night.


\section{Data Reduction}

\subsection{Photometry Reduction}

Reduction of the photometric data was done using the Image Reduction
and Analysis Facility (IRAF).  Bias subtraction and flatfielding of
the VRI frames were accomplished using the ccdred package, and
instrumental magnitudes were obtained using the apphot package.  The
photcal package was then used to perform fits to the extinction and
transformation equations, and to transform the magnitudes to the
standard Cousins system.  An aperture of 3\arcsec\ was used to extract
counts for the program stars, and aperture corrections used to match
the methodology of Landolt (1992).  Errors in the final photometry are
estimated to be 0.03 mag.

\subsection{Extraction of Spectra}

The long-slit spectra were reduced and extracted using IRAF.  Bias levels
were removed by subtracting a median bias frame scaled to match the
overscan signal of each image.  The images were flattened by dividing
by a normalized, median dome flat from which the spectral response of
the illuminating quartz lamp had been removed.  After the removal of
night-sky emission lines, the target spectra were distortion-corrected
and wavelength-calibrated using the consecutively recorded HeAr arc
spectra.  One dimensional spectra were extracted from summed apertures
of 10 to 14 pixels width centered on the spectra.  Correction for
atmospheric extinction was performed using the default IRAF extinction
tables for CTIO, but telluric features (which can be seen in the white
dwarf spectra of Figure 2) were not removed.  Finally, the extracted
spectra were flux calibrated using a recorded spectrum of the
spectrophotometric standard star GJ~440 and the appropriate IRAF flux
table.

\subsection{Assignment of Spectral Types}

We have developed a software program, called ALLSTAR, that matches a
target spectrum to one from a database of 106 standard spectra of K
and M dwarfs previously published by RECONS (see Table~2).  When
expanded to include the complete range of spectral types, ALLSTAR will
likely become the standard algorithm for assigning spectral types in
the NStars Database.

For each target and standard spectrum, ALLSTAR interpolates flux
values at 1~\AA\ intervals between 6500 and 9000~\AA, regardless of
the original spectral resolution.  The spectra are normalized at
7500~\AA, a wavelength which lies in a region that is relatively free
of opacity sources in red dwarfs (and most stars).  To account for
possible spurious normalization of the target spectrum caused by noisy
data, an array of 21 spectra is created by multiplying the normalized
spectrum by integral percentages between 90\% and 110\%.  These 21
spectra are subtracted from each of the standard spectra over the
entire 6500--9000~\AA\ range, and the rms deviation of each difference
spectrum is then computed.  Pixels offset by more than 2$\sigma$ from
the rms deviation are set to zero.  Trial and error has shown that
this threshold effectively rejects unwanted pixels associated with
variable telluric absorption features or detector defects.  If more
than 20\% of the difference spectrum is rejected, then the standard
spectrum from which it was derived is removed from further
consideration.  After rejecting the deviant pixels, ALLSTAR sums the
elements of each difference spectrum and sorts the sums for all
matches in ascending order.  These ordered sums provide a rating
system in which the standard spectrum generating the smallest sum is
the best match to the target spectrum.

The technique embodied in ALLSTAR differs from that used in previous
RECONS efforts (Kirkpatrick, Henry, \& McCarthy 1991) in four ways:
(1) ALLSTAR interpolates fluxes every 1~\AA\ rather than every 3~\AA,
(2) the target spectra are compared with the standard spectra over the
entire 6500--9000~\AA\ range, rather than over pre-assigned chunks of
the spectra, (3) rejection of varying telluric features and detector
defects is carried out in a rigorous, well-defined fashion, and (4)
spectral typing is based on a single, best match to a standard
spectrum rather than a relative ranking of all spectra from bluest to
reddest.  For 100 of our 106 standard stars, ALLSTAR returns the same
spectral types previously reported by RECONS and listed in Table~2.
For the remaining six, ALLSTAR produces spectral types within 0.5
subtype (the published uncertainty) of those previously reported.
These 0.5 subtype discrepancies are attributable to subtle differences
in the reduction techniques, and are not significant.  Therefore, we
have not altered the previously published types.

In Table~2, ``J'' has been appended to some of the previously
published spectral types to denote those stars whose spectra represent
the combined light of multiple components.  The ``J'' (for joint)
designation was not previously used for GJ~22~AC, GJ~352~AB,
GJ~570~BC, G~203-047~AB, GJ~791.2~AB, GJ~829~AB, GJ~896~AC, and
GJ~896~BD because, for many of these systems, close companions have
been discovered since their spectral types were first published.  We
anticipate that other changes in spectral type will occur once spectra
obtained at infrared wavelengths are combined with existing optical
spectra.


\section{Results and Discussion}

Table~3 contains astrometric, photometric, and spectroscopic
information for the 34 stars in our sample.  The photometry is given
on the Johnson ($UBV$) and Cousins ($RI$) systems.  The photometry
from Weis (1996) has been converted to the Cousins system using the
relations of Bessell \& Weis (1987).  Previously reported spectral
types for many of the stars come from Reid {\it et al.} (1995) and
Hawley {\it et al.} (1996), who used narrower-band spectra ($\approx
6200$--7400~\AA) than ours to determine the spectral types of over
2000 known and candidate nearby stars.  Our observations differ from
theirs in that ours cover more than three times the spectral range and
result in somewhat more robust spectral types.  Also listed are
distances to stars with trigonometric parallaxes from Hipparcos
(denoted by H) or from a weighted mean of YPC and Hipparcos
measurements (denoted YH).

Using the $M_V$--spectral type relation of Henry {\it et al.} (1994)
for red dwarfs (ST is spectral type),

\vskip5pt

\noindent\hskip60pt M$_V$ = 0.101 (ST)$^2$ $+$ 0.596 (ST) $+$ 8.96 \hskip100pt (1)
\vskip10pt

\noindent we have estimated distances (last column of Table~3) to
stars for which we have obtained new spectra and that have $V_J$
photometry.  Comparison of the true and predicted distances for the
stars with known trigonometric parallaxes shows that the errors on the
distance estimates are substantial, which is typical of distances
estimated spectroscopically.  However, only for GJ~190 and GJ~433 do
the trigonometric and spectroscopic distances differ by more than a
factor of two.  GJ~433 is a known close binary system (Bernstein 1997;
confirmed by C.\ Leinert, personal communication), and we are
suspicious that GJ~190 may also be a close binary system.

We find that eleven of the candidate nearby stars identified by
Wroblewski and Torres (WT stars) are red dwarfs of types M0.0V to
M5.5V.  The twelfth WT star is a newly identified nearby white dwarf.
The two Calan-ESO stars are also red dwarfs.  As expected, the
Hipparcos stars with large parallax errors are giants and, therefore,
not members of the solar neighborhood.  The four stars in the
photometrically selected sample yield the highest proportion of nearby
stars -- GJ~1123 and GJ~1128 are likely to be within 10~pc, while
GJ~1129 lies just beyond 10~pc.  (We lack the accurate $V_J$
photometry required to estimate the distance to LHS~1957.)  Many of
the targets with estimated distances within 25~pc are being observed
in our southern parallax program, CTIOPI.  Those stars likely to be
within the 10~pc RECONS horizon receive the highest priority.

The spectra of the most interesting stars in our sample are shown in
Figures~1 and 2.  Several noteworthy stars are discussed here, in
alphabetical order.

{\bf ESO 440-064} (spectral type M5.5V) is one of the two latest stars
observed.  It exhibits a prominent H$\alpha$ emission feature.  This
star and {\bf ESO 440-139}, which is estimated to lie at 20~pc, were
revealed during the Calan-ESO effort of Ruiz and collaborators (Ruiz
{\it et al.} 1993) to reveal new proper motion stars in the southern
sky.

{\bf GJ 432 B} is a 15th magnitude companion $17''$ from a 6th
magnitude K0 dwarf.  Observing difficulty precludes much information
for this star.  Its spectrum is similar to our white dwarf
spectrophotometric standard, GJ~440.  We cannot determine with our
spectral coverage whether the star has type DC or DQ, but it is not
type ``m'' as reported by Gliese \& Jahreiss (1991).

{\bf GJ 1123, GJ 1128, GJ 1129} and {\bf WT 84} are four stars with
estimated distances within 15~pc.  All are being observed as
high-priority CTIOPI targets because they may lie within the RECONS
horizon of 10~pc.  WT~84 (spectral type M5.5V) is one of the two
latest stars observed and exhibits a prominent H$\alpha$ emission
feature.

{\bf GJ 2036 A and B} have enormous H$\alpha$ emission features.

{\bf HIP 14555} (=~LTT 1479 =~GJ 1054A) was observed instead of the
intended target, HIP~14559, which lies 28\arcsec\ to the east.  The
Hipparcos parallax of HIP~14559 ($V_J = 11.72$), 0\arcpt11473 $\pm$
0\arcpt03398, has a large error because of the proximity of HIP~14555
($V_J = 10.21$).  HIP~14555 has spectral type M1.0V and a prominent
H$\alpha$ emission feature.  Using our new spectral type and the $V_J$
magnitude from Hipparcos, we estimate a distance to HIP~14555 of
12.9~pc, which is $5\sigma$ less than the distance obtained from the
Hipparcos parallax of 0\arcpt05238 $\pm$ 0\arcpt00503.  The fainter
star found 64\arcsec\ to the southwest is LTT~1477 (=~GJ~1054B), which
has common proper motion with HIP~14555.

{\bf HIP 15689} ($V_J = 12.16$) lies 24\arcsec\ southwest of HIP 15690
($V_J = 8.83$).  The Hipparcos parallax of 0\arcpt22745 $\pm$
0\arcpt06179 for HIP~15689 has a large error because of the proximity
of HIP~15690.  Our spectrum of HIP~15689 lacks a K~I feature, has a
weak Na~I feature, and has a strong Ca~II triplet.  These features
clearly indicate that the star is a giant or supergiant.  It is
therefore not a nearby star.

{\bf HIP 20968} ($V_J = 11.42$) lies 21\arcsec\ northeast of {\bf HIP~20965}
($V_J = 7.73$).  The Hipparcos parallax of 0\arcpt12070 $\pm$
0\arcpt05647 for HIP~20968 has a large error because of the proximity
of HIP~20965 (parallax 0\arcpt00218 $\pm$ 0\arcpt00189).  Our spectra
indicate that both stars are giants or supergiants.  HIP~20968 is
therefore not a nearby star.  HIP~20965's spectrum shows the CN band
at 7900 \AA.

{\bf WT 248} (M3.0V) is estimated to lie at a distance of 37 pc,
despite its large proper motion (1\arcpt197~yr$^{-1}$).  Its spectrum
does not show the obvious CaH band around 6900 \AA\ that is
characteristic of subdwarfs, as might be expected for such a high
velocity star.

{\bf WT 1759} is a newly identified nearby white dwarf.  Its spectrum
is virtually identical to that of our spectroscopic standard, GJ 440,
which is a DQ6 white dwarf with a temperature of $\sim$8500~K
(Bergeron {\it et al.} 2001).  With our spectral coverage, we cannot
determine if the star has type DC or DQ.  Assuming that WT~1759 has
the same absolute magnitude as GJ~440, we estimate that the distance
to WT~1759 is 28~pc.

\section{Concluding Remarks}

The recent identifications of candidate nearby stars from the proper
motion studies of Wroblewski {\it et al.}, Ruiz {\it et al.}, and
others, and from photometric sky surveys such as DENIS, 2MASS and
SDSS, suggest that many nearby stars remain undiscovered.  In essence,
this paper represents a small step in fingerprinting some suspected
nearby stars via spectroscopy.  We have established a method for
consistent spectral typing that will provide definitive types both for
the RECONS effort (horizon 10~pc) and the more extensive NASA/NSF
NStars Project (horizon 25~pc).  Using this method, we report the
first spectral types on a standard system for 16 nearby star
candidates.  We also provide updated spectral types for 18 other stars
using broader spectral coverage than was previously available.

This work will allow us to improve the luminosities, colors, and
temperatures for the ubiquitous red dwarfs, as well as broaden the
database used to investigate the luminosity function, mass function,
kinematics, and multiplicity of stars in the solar neighborhood.  The
nearest objects, such as GJ~1123 and GJ~1128 from this study, will be
prime targets of upcoming NASA missions like SIRTF, SIM, and TPF, as
well as being additions to the target lists of SETI efforts like
Project Phoenix.

\acknowledgments

TJH and TCB thank the Space Telescope Science Institute in Baltimore,
Maryland for providing the facilities and resources necessary to
conduct this research during the Summer Student Programs of 1996 and
1997.  TJH and LMW thank the NStars Project for current support, and
are grateful for the dedicated work of Dana Backman, Jerry Blackwell,
and Takeshi Okimura who have made the NStars Database a reality.  We
also would like to thank members of the CTIOPI team, specifically Phil
Ianna, Ricky Patterson, and John Subasavage, for their work on the
CTIOPI photometry.  More information on nearby star work can be found
at the RECONS (\url{http://www.chara.gsu.edu/RECONS}) and NStars
(\url{http://nstars.arc.nasa.gov}) websites.

\clearpage


\figcaption[fig1.eps]{Spectra of some of the nearest suspects in our
sample of candidate nearby stars.  The spectra and photometry of
GJ~1129, GJ~1123, GJ~1128, and WT~84 suggest that these stars lie
within 15~pc (see Table~3).  The spectra of GJ~2036~A and B exhibit
enormous H$\alpha$ emission features.  Important spectral features are
labelled at the top.  The absorption complex at 9300 \AA~and redward
is due primarily to H$_2$O in the Earth's atmosphere.  \label{fig1}}

\figcaption[fig2.eps]{Spectra of the three white dwarfs and the three
giants.  WT~1759 is a newly identified white dwarf with an estimated
distance of 28~pc.  GJ~432~B is a nearby white dwarf companion to a K0
dwarf.  The spectra of both white dwarfs are similar to that of the
DQ6 spectrophotometric standard, GJ~440.  HIP~15689, HIP~20968, and
HIP~20965 are giants or supergiants, as indicated by the lack of K~I
absorption near 7700 \AA, the weak NaI absorption near 8200 \AA, and
the strong CaII triplet in the 8500--8700 \AA\ window.  Note also the
CN absorption feature near 7900 \AA\ in the spectrum of HIP~20965.
Important spectral features are labelled at the top. The absorption
complex at 9300 \AA~and redward is due primarily to H$_2$O in the
Earth's atmosphere.  \label{fig2}}


\clearpage

\begin{deluxetable}{cccc}

\tabletypesize{\scriptsize}
\tablecaption{Number of Stellar Systems within 25 Parsecs\tablenotemark{a}. \label{tbl-1}}
\tablewidth{0pt}
\tablehead{
\colhead{Region}& \colhead{\# Systems}&  \colhead{Total}&     \colhead{Fraction} \\
\colhead{of Sky}& \colhead{Known}     &  \colhead{Predicted}& \colhead{Missing} \\
}                                       
\startdata                              
$+$90 to $+$30  &        575&                  1375&                 58\% \\
$+$30 to $+$00  &        578&                  1375&                 58\% \\
$-$00 to $-$30  &        463&                  1375&                 66\% \\
$-$30 to $-$90  &        395&                  1375&                 71\% \\
\\                                                  
TOTAL           &       2011&                  5500&                 63\% \\
\\
\enddata

\tablenotetext{a}{in NStars Database as of 01 July 2001}

\end{deluxetable}

\clearpage


\clearpage

\begin{deluxetable}{llcccc}

\tabletypesize{\scriptsize}
\tablecaption{List of Spectral Standards. \label{tbl-2}}
\tablewidth{0pt}
\tablehead{
\colhead{Name}& \colhead{Comp.}& \colhead{RA}& \colhead{Dec}& \colhead{Spectral}& \colhead{Ref.}\\
\colhead{}& \colhead{}& \colhead{(J2000)}& \colhead{(J2000)}& \colhead{type}& \colhead{}\\
}
\startdata
GJ 1002		&\nodata&00 06 43.8& $-$07 32 22		&M5.5V 	& Hen94  \\  
GJ 1005    	&AB    	&00 15 28.1& $-$16 08 02 	&M4.0VJ	& Hen94  \\  
GJ   15    	&A     	&00 18 22.9& $+$44 01 23		&M1.5V 	& Hen94  \\  
GJ   15    	&B     	&00 18 22.9& $+$44 01 23 	&M3.5V 	& Hen94  \\  
GJ 2005    	&ABCD  	&00 24 42.0& $-$27 08 52 	&M5.5VJ	& Hen99  \\  
GJ   22    	&AC	&00 32 26.0& $+$67 14 00 	&M2.0VJ	& McC91  \\
GJ   22    	&B      &00 32 26.0& $+$67 14 00 	&M3.0V 	& McC91  \\
GJ   34    	&B   	&00 49 06.3& $+$57 48 55 	&K7.0V 	& Hen94  \\  
GJ   51		&\nodata&01 03 18.0& $+$62 22 00 	&M5.0V 	& Kir91  \\
GJ   54.1       &\nodata&01 12 30.6& $-$16 59 57 	&M4.5V 	& Hen94  \\  
GJ   65    	&A    	&01 39 01.3& $-$17 57 01 	&M5.5V 	& Kir91  \\  
GJ   65    	&B    	&01 39 01.3& $-$17 57 01 	&M6.0V 	& Kir91  \\  
GJ   83.1       &\nodata&02 00 13.2& $+$13 03 08 	&M4.5V 	& Kir91  \\  
GJ  105    	&B    	&02 36 16.0& $+$06 52 12		&M3.5V 	& Hen94  \\  
GJ  109         &\nodata&02 44 15.5& $+$25 31 24 	&M3.0V 	& Hen94  \\  
GJ 1061         &\nodata&03 36 00.0& $-$44 30 46 	&M5.5V 	& Hen97  \\  
LP 944-020      &\nodata&03 39 35.2& $-$35 25 41         &$>$M9.0V&Kir97  \\  
GJ  185    	&AB   	&05 02 28.4& $-$21 15 24 	&K7.0VJ	& Hen94  \\  
GJ  205         &\nodata&05 31 27.4& $-$03 40 38 	&M1.5V 	& Kir91  \\  
GJ  213         &\nodata&05 42 09.3& $+$12 29 22 	&M4.0V 	& Kir91  \\  
G 099-049       &\nodata&06 00 03.6& $+$02 42 20 	&M3.5V 	& Hen94  \\  
LHS 1805        &\nodata&06 01 09.7& $+$59 35 54 	&M3.5V 	& Hen94  \\  
GJ  226         &\nodata&06 10 19.8& $+$82 06 24 	&M2.5V 	& Kir91  \\  
GJ  229    	&A    	&06 10 34.6& $-$21 51 53 	&M1.0V 	& Kir91  \\  
GJ  232         &\nodata&06 24 41.6& $+$23 25 59 	&M4.0V 	& Kir91  \\  
GJ  234    	&AB   	&06 29 23.4& $-$02 48 50 	&M4.5VJ	& Kir91  \\  
GJ  250    	&B    	&06 52 18.1& $-$05 11 25 	&M2.5V 	& Kir91  \\  
GJ  251         &\nodata&06 54 49.0& $+$33 16 05 	&M3.0V 	& Kir91  \\  
GJ 1093         &\nodata&06 59 28.4& $+$19 20 52 	&M5.0V 	& Hen94  \\  
GJ  268    	&AB   	&07 10 01.8& $+$38 31 46 	&M4.5VJ	& Kir91  \\  
GJ  273         &\nodata&07 27 24.5& $+$05 13 33 	&M3.5V 	& Kir91  \\  
GJ  283    	&B    	&07 40 20.7& $-$17 24 52 	&M6.0V 	& Hen94  \\  
GJ  285         &\nodata&07 44 40.2& $+$03 33 09 	&M4.0V 	& Hen94  \\  
GJ  299         &\nodata&08 11 57.5& $+$08 46 28 	&M4.0V 	& Hen94  \\  
GJ  300         &\nodata&08 12 40.8& $-$21 33 10 	&M3.5V 	& Hen94  \\  
GJ 1111         &\nodata&08 29 49.5& $+$26 46 37 	&M6.5V 	& Hen94  \\  
LHS 2065        &\nodata&08 53 36.0& $-$03 29 28 	&M9.0V 	& Kir91  \\  
GJ 1116    	&AB   	&08 58 14.9& $+$19 45 43 	&M5.5VJ	& Hen94  \\  
GJ  338    	&A    	&09 14 22.8& $+$52 41 12 	&M0.0V 	& Kir91  \\  
GJ  338    	&B    	&09 14 24.7& $+$52 41 11 	&K7.0V 	& Kir91  \\
GJ  352    	&AB     &09 31 19.4& $-$13 29 19		&M3.0VJ	& Kir91  \\
GJ  380         &\nodata&10 11 22.1& $+$49 27 15 	&K7.0V 	& Kir91  \\  
GJ  381  	&\nodata&10 12 05.0& $-$02 41 12 	&M2.5V 	& Kir91  \\
GJ  382         &\nodata&10 12 17.7& $-$03 44 44 	&M2.0V 	& Kir91  \\  
GJ  393         &\nodata&10 28 55.5& $+$00 50 28 	&M2.0V 	& Hen94  \\  
LHS  292        &\nodata&10 48 12.6& $-$11 20 14 	&M6.5V 	& Hen94  \\  
GJ  402         &\nodata&10 50 52.1& $+$06 48 29 	&M4.0V 	& Kir91  \\  
GJ  406         &\nodata&10 56 29.2& $+$07 00 53 	&M6.0V 	& Kir91  \\  
GJ  411         &\nodata&11 03 20.2& $+$35 58 12 	&M2.0V 	& Kir91  \\  
GJ  412    	&A    	&11 05 28.6& $+$43 31 36 	&M1.0V 	& Hen94  \\  
GJ  412    	&B    	&11 05 30.4& $+$43 31 18 	&M5.5V 	& Hen94  \\
GJ  436   	&\nodata&11 42 09.0& $+$26 42 24 	&M3.0V 	& Kir91  \\
GJ  445         &\nodata&11 47 41.4& $+$78 41 28 	&M3.5V 	& Hen94  \\  
GJ  447         &\nodata&11 47 44.4& $+$00 48 16 	&M4.0V 	& Hen94  \\  
GJ 1156         &\nodata&12 19 00.3& $+$11 07 31 	&M5.0V 	& Hen94  \\  
GJ  473    	&AB   	&12 33 17.2& $+$09 01 15 	&M5.5VJ	& Hen92  \\  
GJ  514         &\nodata&13 29 59.8& $+$10 22 38 	&M1.0V 	& Hen94  \\  
GJ  526         &\nodata&13 45 43.8& $+$14 53 29 	&M1.5V 	& Hen94  \\  
GJ  551         &\nodata&14 29 43.0& $-$62 40 46 	&M5.5V 	& Hen97  \\  
GJ  555         &\nodata&14 34 16.8& $-$12 31 10 	&M3.5V 	& Hen94  \\  
LHS 3003        &\nodata&14 56 38.5& $-$28 09 51 	&M7.0V 	& Kir95  \\  
GJ  570    	&BC   	&14 57 26.5& $-$21 24 41 	&M1.0VJ	& Hen94  \\  
TVLM 513-46546  &\nodata&15 01 07.9& $+$22 50 02 	&M8.5V 	& Kir95  \\  
GJ  581         &\nodata&15 19 26.8& $-$07 43 20 	&M2.5V 	& Hen94  \\  
GJ  623    	&AB   	&16 24 09.3& $+$48 21 10 	&M2.5VJ	& Hen94  \\  
GJ  625         &\nodata&16 25 24.6& $+$54 18 15 	&M1.5V 	& Hen94  \\  
GJ  628         &\nodata&16 30 18.1& $-$12 39 45 	&M3.0V 	& Hen94  \\  
GJ  643         &\nodata&16 55 25.2& $-$08 19 21 	&M3.5V 	& Kir91  \\  
GJ  644    	&ABD  	&16 55 28.8& $-$08 20 11 	&M2.5VJ	& Hen94  \\  
GJ  644    	&C    	&16 55 35.8& $-$08 23 40 	&M7.0V 	& Kir91  \\  
G 203-047  	&AB   	&17 09 31.5& $+$43 40 53 	&M3.5VJ	& Hen94  \\  
GJ  661    	&AB   	&17 12 07.9& $+$45 39 57 	&M3.0VJ	& Hen94  \\  
GJ  673         &\nodata&17 25 45.2& $+$02 06 41 	&K7.0V 	& Hen94  \\  
GJ  686         &\nodata&17 37 53.4& $+$18 35 30 	&M0.0V 	& Hen94  \\  
GJ  687         &\nodata&17 36 25.9& $+$68 20 21 	&M3.0V 	& Hen94  \\  
GJ  699         &\nodata&17 57 48.5& $+$04 41 36 	&M4.0V 	& Kir91  \\  
GJ  701         &\nodata&18 05 07.6& $-$03 01 53 	&M0.0V 	& Hen94  \\  
GJ 1224         &\nodata&18 07 32.9& $-$15 57 51 	&M4.5V 	& Hen94  \\  
LHS 3376        &\nodata&18 18 57.7& $+$66 11 32 	&M4.5V 	& Hen94  \\  
GJ 1230    	&AC   	&18 41 09.2& $+$24 47 08 	&M4.5VJ	& Hen94  \\  
GJ 1230    	&B     	&18 41 09.2& $+$24 47 15 	&M4.5V 	& Hen94  \\  
GJ  725    	&A    	&18 42 46.7& $+$59 37 49 	&M3.0V 	& Kir91  \\  
GJ  725    	&B    	&18 42 46.9& $+$59 37 37 	&M3.5V 	& Kir91  \\  
GJ  729         &\nodata&18 49 49.4& $-$23 50 10 	&M3.5V 	& Hen94  \\  
GJ  752    	&A    	&19 16 55.3& $+$05 10 08 	&M3.0V 	& Kir91  \\  
GJ  752    	&B    	&19 16 58.3& $+$05 09 01 	&M8.0V 	& Kir91  \\  
GJ 1245    	&AC   	&19 53 54.2& $+$44 24 55 	&M5.5VJ	& Kir91  \\  
GJ 1245    	&B    	&19 53 55.2& $+$44 24 56 	&M6.0V 	& Kir91  \\  
GJ  791.2  	&AB   	&20 29 48.0& $+$09 41 23 	&M4.5VJ	& Kir91  \\  
GJ  809         &\nodata&20 53 19.8& $+$62 09 16 	&M0.0V 	& Hen94  \\  
GJ  820    	&A    	&21 06 53.9& $+$38 44 58 	&K5.0V 	& Kir91  \\  
GJ  820    	&B    	&21 06 55.3& $+$38 44 31 	&K7.0V 	& Kir91  \\  
GJ  829    	&AB   	&21 29 36.8& $+$17 38 36 	&M3.5VJ	& Hen94  \\
GJ  831    	&ABC  	&21 31 18.9& $-$09 47 22 	&M4.5VJ	& Hen94  \\  
GJ  846         &\nodata&22 02 09.0& $+$01 23 54		&M0.5V 	& Kir91  \\
GJ  860    	&A   	&22 27 59.5& $+$57 41 45 	&M3.0V 	& Hen94  \\  
GJ  860    	&B   	&22 27 59.5& $+$57 41 45 	&M4.0V 	& Hen94  \\  
GJ  866    	&ABC  	&22 38 33.4& $-$15 18 07 	&M5.0VJ	& Kir91  \\  
GJ  873         &\nodata&22 46 49.7& $+$44 20 02 	&M3.5V 	& Hen94  \\  
GJ  876      &Ap\tablenotemark{a}&22 53 16.7& $-$14 15 49&M3.5VJ	& Hen94  \\  
GJ  880         &\nodata&22 56 34.8& $+$16 33 12 	&M1.5V 	& Hen94  \\  
GJ  896    	&AC   	&23 31 52.2& $+$19 56 14 	&M3.5VJ	& Hen94  \\
GJ  896    	&BD   	&23 31 52.2& $+$19 56 14 	&M4.5VJ	& Hen94  \\
GJ 1286         &\nodata&23 35 10.7& $-$02 23 25 	&M5.5V 	& Hen94  \\  
GJ  905         &\nodata&23 41 54.7& $+$44 10 30 	&M5.5V 	& Hen94  \\  
GJ  908         &\nodata&23 49 12.5& $+$02 24 04 	&M1.0V 	& Hen94  \\	
\enddata
\tablenotetext{a}{p indicates probable planetary companion(s)}

\tablenotetext{~}{references: Hen92, Hen94, Hen97, Hen99 = Henry {\it
et al.}~1992, 1994, 1997, 1999; Kir91, Kir95, Kir97 = Kirkpatrick {\it
et al.}~1991, 1995, 1997, McC91 = McCarthy {\it et al.}~1991}

\end{deluxetable}

\clearpage


\doublespace
\voffset50pt{
{
\begin{deluxetable}{lcccrrrrrlcccccc}
\rotate
\tabletypesize{\scriptsize}
\tablecaption{Sample stars\label{tbl-3}}
\tablewidth{0pt}

\tablehead{
\colhead{Name}& \colhead{RA}& \colhead{Dec}& \colhead{$\mu$ \hskip10pt $\theta$}& \colhead{$U_J$}& 
\colhead{$B_J$}& \colhead{$V_J$}& \colhead{$R_C$}& \colhead{$I_C$}& \colhead{Ref.}& \colhead{Previous}& \colhead{Ref.}& \colhead{Adopted}& \colhead{Exposure}& \colhead{Known}& \colhead{Estimated} \\

\colhead{}& \colhead{(J2000)}& \colhead{(J2000)}& \colhead{}& 
\colhead{}& \colhead{}&\colhead{}& \colhead{}& \colhead{}& \colhead{}& \colhead{Spec Type}& \colhead{}& \colhead{Spec Type}& \colhead{(sec)}& \colhead{dist (pc)}& \colhead{dist (pc)}
}

\startdata
\vspace{-25pt} \\
\multicolumn{16}{c}{Stars with trigonometric parallaxes} \\
\tableline \vspace{-15pt} \\
HIP 14555  &03 07 55.7& $-$28 13 11& 0.360 250.6  & \nodata& \nodata& 10.21  & \nodata& \nodata& HIP97          & \nodata& \nodata& M1.0V & 90      &  19.1 H &   12.9   \\
HIP 20965  &04 29 43.4& $-$29 01 47& 0.045 223.8  & \nodata& \nodata&  7.73  & \nodata& \nodata& HIP97          &KIII/IV & SIMBAD & giant+& 10      &$>$100 H &  \nodata \\
GJ 2036 A  &04 53 31.2& $-$55 51 37& 0.130 059.0  & 13.78  & 12.70  & 11.13  & \nodata& \nodata& GJ79           & M2.0V  & Haw96  & M3.0V & 60      &  11.2 H &    7.8   \\
GJ 2036 B  &04 53 31.2& $-$55 51 37& 0.130 059.0  & 15.03  & 13.75  & 12.15  & \nodata& \nodata& GJ79           & M3.5V  & Haw96  & M4.0V & 60      &  11.2 H &    6.9   \\
LHS 1731   &05 03 20.1& $-$17 22 25& 0.512 209.0  & \nodata& 13.32  & 11.69  & 10.56  &  9.15  & Wei96          & M3.0V  & Rei95  & M3.0V & 1560    &   9.3 H &   10.1   \\
GJ 190 	   &05 08 35.0& $-$18 10 19& 1.376 156.6  & 12.86  & 11.84  & 10.32  &  9.17  &  7.67  & Bes90,Leg92    & M3.5V  & Rei95  & M3.5V & 420     &   9.3 YH&    4.1   \\
GJ 203	   &05 28 00.2& $+$09 38 38& 0.783 195.1  & 15.33  & 14.12  & 12.49  & 11.29  &  9.80  & Bes90,GJ91     & M3.5V  & Rei95  & M3.5V & 1200,135&   9.7 YH&   10.8   \\
GJ 239	   &06 37 10.8& $+$17 33 53& 0.885 293.4  & 12.27  & 11.14  &  9.64  &  8.72  &  7.75  & Wei96,Leg92    & K7.0V  & Rei95  & M0.0V & 300,35  &   9.8 YH&   13.7   \\
GJ 1125	   &09 30 44.6& $+$00 19 22& 0.760 229.0  & 14.52  & 13.27  & 11.72  & 10.58  &  9.13  & Wei96,GJ91     & M3.5V  & Rei95  & M3.0V & 30      &   9.9 YH&   10.3   \\
GJ 358	   &09 39 46.4& $-$41 04 03& 0.663 305.0  & 13.32  & 12.21  & 10.72  &  9.64  &  8.29  & Leg92          & M2.0V  & Haw96  & M3.0V & 600,22  &   9.5 YH&    6.5   \\
GJ 432 B   &11 34 30.5& $-$32 50 01& 1.063 320.3  & \nodata& \nodata& \nodata& \nodata& \nodata&                & DC     & SIMBAD & DC/DQ & 600     &   9.5 YH&  \nodata \\
GJ 433 AB  &11 35 26.9& $-$32 32 24& 0.780 186.0  & 12.51  & 11.36  &  9.84  &  8.84  &  7.69  & Bes90,GJ91     & M1.5V  & Haw96  & M2.0VJ& 420,20  &   9.1 YH&    4.1   \\
GJ 442 B   &11 46 32.5& $-$40 29 47& 1.592 284.4  & \nodata& \nodata& \nodata& \nodata& \nodata&                & M4.0V  & Haw96  & M4.0V & 480     &   9.2 YH&  \nodata \\
GJ 480.1   &12 40 46.3& $-$43 33 59& 1.047 311.7  & 15.36  & 13.97  & 12.24  & 11.07  &  9.63  & Bes90,GJ91     & M3.0V  & Haw96  & M3.0V & 120     &   7.8 YH&   13.1   \\
\tableline \vspace{-15pt} \\
\multicolumn{16}{c}{Candidate nearby stars} \\
\tableline \vspace{-15pt} \\
WT 60	   &01 51 59.7& $-$57 47 58& 0.652 212.2  & \nodata& \nodata& \nodata& \nodata& \nodata&                & \nodata& \nodata& M4.0V & 600     &  \nodata&  \nodata \\
WT 84	   &02 17 27.9& $-$59 22 43& 0.559 212.6  & \nodata& \nodata& 15.88  & 14.29  & 12.28  & Pat98          & \nodata& \nodata& M5.5V & 600     &  \nodata&   13.1   \\
WT 1356	   &03 13 19.7& $-$16 38 47& 0.682 235.4  & \nodata& \nodata& \nodata& \nodata& \nodata&                & \nodata& \nodata& M4.5V & 600     &  \nodata&  \nodata \\
HIP 15689  &03 22 05.5& $-$13 16 44& unknown\tablenotemark{a}& 12.08& 12.04  & 11.54  & \nodata& \nodata& Sin96 & \nodata& \nodata& giant+& 140     &  \nodata&  \nodata \\
WT 133	   &04 02 13.9& $-$43 25 26& 0.561 175.5  & \nodata& \nodata& 16.09  & 14.71  & 13.00  & this paper     & \nodata& \nodata& M4.5V & 600     &  \nodata&   30.2   \\
WT 135	   &04 11 27.1& $-$44 18 09& 0.689 066.7  & \nodata& \nodata& \nodata& \nodata& \nodata&                & \nodata& \nodata& M2.5V & 420     &  \nodata&  \nodata \\
HIP 20968  &04 29 44.9& $-$29 01 37& unknown\tablenotemark{a}& \nodata& \nodata& 11.42& \nodata& \nodata& HIP97 & \nodata& \nodata& giant+& 60      &  \nodata&  \nodata \\
WT 207	   &07 02 36.6& $-$40 06 29& 0.624 105.3  & \nodata& \nodata& 15.08  & 13.91  & 12.31  & Pat98          & \nodata& \nodata& M4.0V & 600     &  \nodata&   26.5   \\
WT 214	   &07 28 40.1& $-$61 20 41& 0.626 319.5  & \nodata& \nodata& 16.06  & 14.80  & 13.17  & Pat98          & \nodata& \nodata& M4.0V & 600     &  \nodata&   41.7   \\
WT 233	   &07 56 13.7& $-$67 05 19& 0.792 325.2  & \nodata& \nodata& 16.23  & 15.34  & 14.44  & this paper     & \nodata& \nodata& M0.0V & 600     &  \nodata&  284.4   \\
LHS 1957   &07 57 00.2& $-$45 37 23& 0.666 341.6  & \nodata& \nodata& \nodata& \nodata& \nodata&                & \nodata& \nodata& M2.5V & 270     &  \nodata&  \nodata \\
GJ 1123	   &09 16 45.0& $-$77 49 42& 1.023 139.3  & 15.89  & 14.74  & 13.10  & \nodata& \nodata& GJ91           & M4.5V  & Haw96  & M4.5V & 210     &  \nodata&    7.6   \\
GJ 1128	   &09 42 53.0& $-$68 54 06& 1.057 348.0  & 15.73  & 14.51  & 12.78  & \nodata& \nodata& GJ91           & M4.5V  & Haw96  & M4.5V & 120     &  \nodata&    6.6   \\
WT 244	   &09 44 28.6& $-$73 58 39& 0.524 256.9  & \nodata& \nodata& 15.24  & 13.85  & 12.07  & Pat98          & \nodata& \nodata& M4.5V & 600     &  \nodata&   20.4   \\
GJ 1129	   &09 44 48.0& $-$18 12 48& 1.633 264.0  & 15.47  & 14.19  & 12.60  & \nodata& \nodata& GJ91           & M4.0V  & Rei95  & M3.5V & 120     &  \nodata&   11.6   \\
WT 248	   &10 05 54.9& $-$67 21 31& 1.197 264.5  & \nodata& \nodata& 14.51  & 13.43  & 12.02  & Pat98          & \nodata& \nodata& M3.0V & 600     &  \nodata&   37.2   \\
WT 1759	   &10 12 01.9& $-$18 43 34& 0.508 264.8  & \nodata& \nodata& 15.44  & 15.20  & 14.97  & this paper     & \nodata& \nodata& DC/DQ & 600     &  \nodata&   28.3\tablenotemark{b}\\
WT 1760	   &10 12 06.2& $-$28 51 38& 0.505 144.8  & \nodata& \nodata& 16.19  & 14.84  & 13.14  & this paper     & \nodata& \nodata& M4.0V & 600     &  \nodata&   44.3   \\
ESO 440-064 &11 48 48.5& $-$28 33 27& 0.710 260.0  & \nodata& \nodata& \nodata& 14.88  & 12.97  & this paper     & M6.4V  & Rui95  & M5.5V & 600     &  \nodata&  \nodata \\
ESO 440-139 &12 03 27.5& $-$29 22 49& 0.310 316.0  & \nodata& \nodata& 15.19  & \nodata& \nodata& Rui95          & M5.5V  & Rui95  & M4.5V & 600     &  \nodata&   20.0   \\

\enddata

\tablenotetext{a}{The Hipparcos proper motions have enormous errors,
probably caused by the effects of nearby bright stars.}
\tablenotetext{b}{Distance derived by comparison to the white dwarf
standard, GJ 440.}
\tablenotetext{~}{references: Bes90 = Bessell 1990; GJ79, GJ91 =
Gliese \& Jahrei$\ss$ 1979, 1991; Haw96 = Hawley {\it et al.}~1996;
HIP97 = ESA 1997; Leg92 = Leggett 1992; Pat98 = Patterson {\it et
al.}~1998; Rei95 = Reid {\it et al.}~1995; Rui95 = Ruiz {\it et
al.}~1995; SIMBAD = SIMBAD database, operated at CDS, Strasbourg,
France; Sin96 = Sinachopoulos \& van Dessel 1996; Wei96 = Weis 1996}

\end{deluxetable}
}
}

\vskip-100pt
\plotone{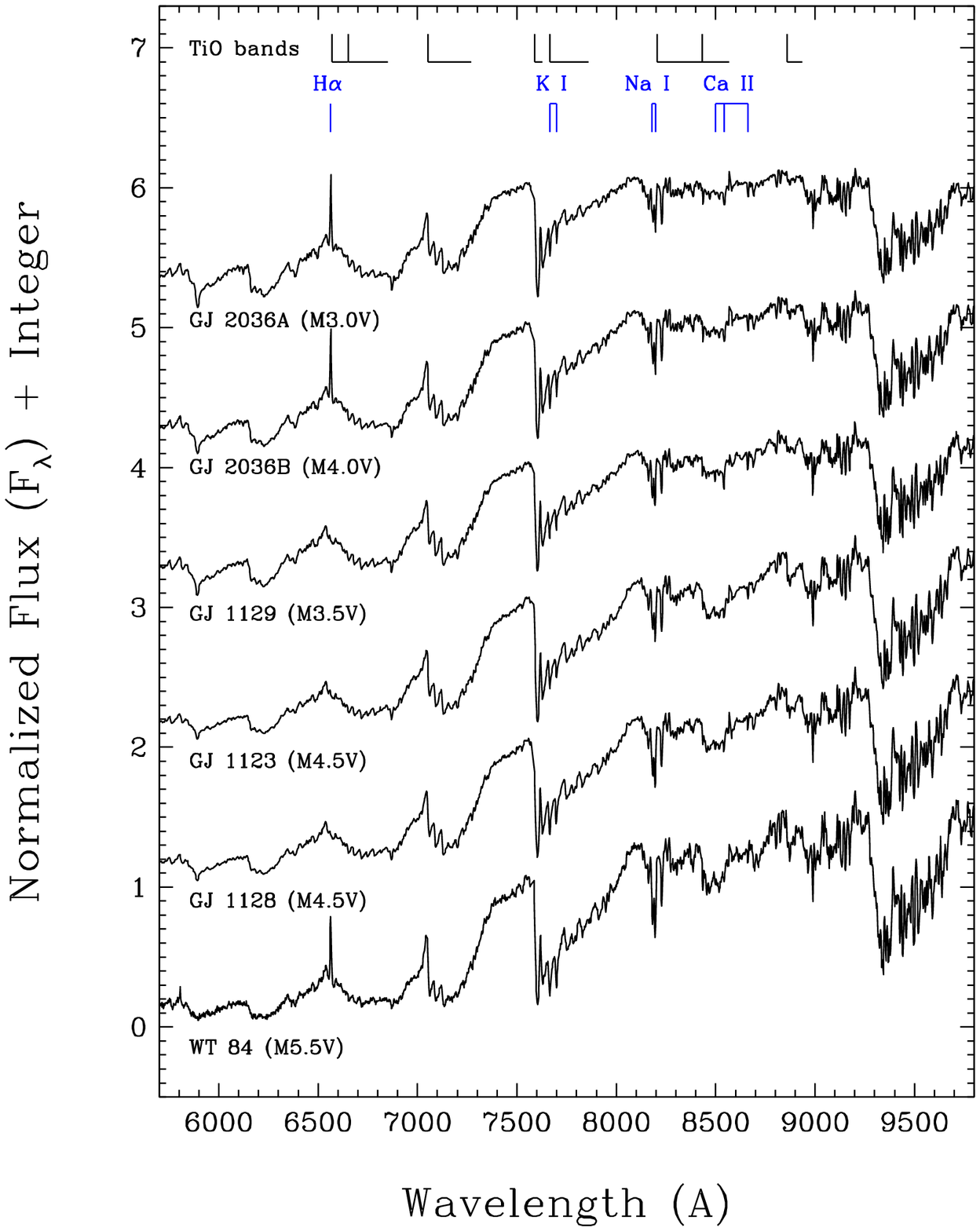}

\vskip-100pt
\plotone{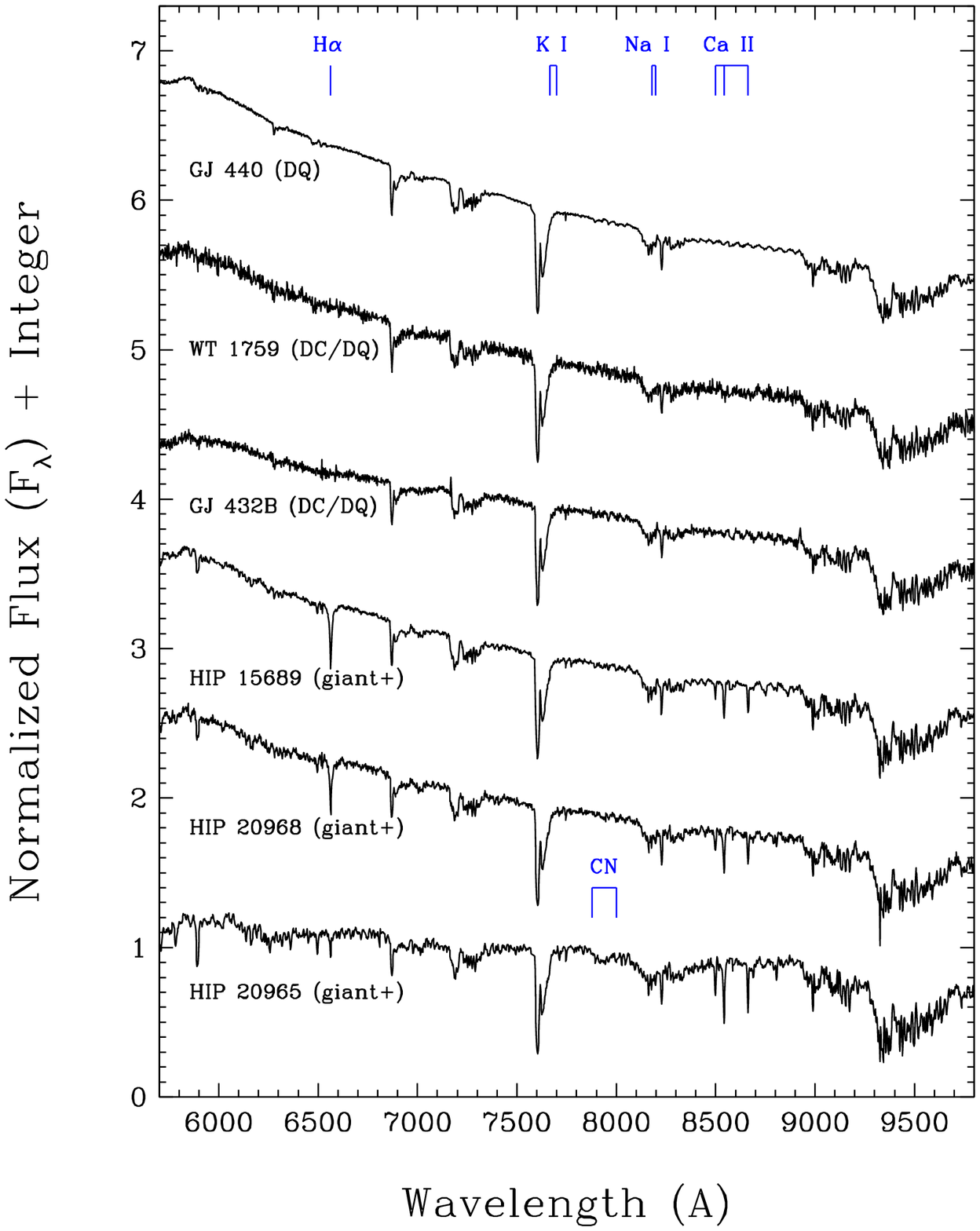}

\end{document}